\documentclass[12pt]{article}
 \usepackage{epsfig}
 \usepackage{cite} \usepackage{amssymb}
 \usepackage{a4wide}
 \usepackage{comment}
\def\beq{\begin{equation}}
\def\eeq{\end{equation}}
 \newcommand{\form}[1]{(\ref{#1})}
\begin{document}
  \bibliographystyle{unsrt}


\begin{centering}
 \begin{flushright}
hep-ph/0605326\\
 CERN-PH-TH/2006-101 \\
 October 2006
 \end{flushright}
 \vspace{0.1in}
 {\Large {\bf 
 Probing Brane-World Scenarios with Vacuum \\
 \smallskip
 \smallskip
 Refraction  of Light Using Gamma-Ray Bursts  
}}

\vspace{0.3in}

{
 {\bf Merab Gogberashvili},$^{a}$~ {\bf Alexander~S. Sakharov},$^{b}$~ 
 {\bf Edward~K.G. Sarkisyan}$^{c}$
\bigskip
\bigskip

{\footnotesize 
$^a$ Andronikashvili Institute of Physics, 6 Tamarashvili
St., Tbilisi 0177, Georgia \\
$^b$ Theory Division, Physics Department, CERN, 1211 Geneva 23,
Switzerland, and \\
Swiss Institute of Technology, ETH-Zurich, 8093 Zurich,
Switzerland \\
$^c$ EP Division, Physics Department, CERN, 1211 Geneva 23,
Switzerland, and \\
School of Physics and Astronomy, The University of Manchester, Manchester 
M13 9PL, UK\\}
}
\vspace{0.2in}

\vspace{0.4in}
\end{centering}
\begin{center}
{\bf Abstract}
\end{center}
\vspace{0.2in}

{\small\noindent
\noindent
We argue that in fat brane-world scenarios 
 the 
light propagating in vacuum
 will,
 because of massive ``Kaluza--Klein''  (KK) excitations, 
  experience a refraction. The
motion of a photon inside a fat brane can be decomposed in the 
longitudinal
and transverse directions with respect to the surface of the brane. 
 Since the
light observable propagation is
related only with the longitudinal motion, the observed speed of light
 depends on the  value of the momentum transverse fraction 
contributing as 
the massive KK excitations. 
This is directly connected with 
 the energy of the particles emitting the light, and hence with the
frequency of the light itself. Using  recent results on the
arrival times of radiation of different
energies from the measurements of 
 gamma-ray bursters with known redshifts, we 
establish
the limit $M>620$~TeV on the inverse thickness of the brane,  and thus on 
the masses of
the KK excitations. This limit exceeds by at least  one order of 
magnitude the 
typical
energy
scale currently in use to characterize brane phenomena in the realm 
of 
future colliders.}

\vspace{0.5in}
\begin{flushleft}
CERN-PH-TH/2006-101 \\
October 2006
\end{flushleft}

\vspace{0.25in}
\begin{flushleft}
\end{flushleft}

\newpage

\section{Introduction}

The constancy of the velocity of light is one of the tenets of  modern
physics. In particular, the special and general theories of relativity
postulate a single universal velocity of light.  However, there is a
general expectation of a need, in many theoretical efforts to find a
synthesis of general relativity and quantum mechanics~\cite{qgrav}, for
great sophistication in the discussion of the propagation of light in 
vacuum.

There are  several theoretical models that consider the modification of
electrodynamics at high energies. A hypothetical Lorentz symmetry
deformation, considered as an explanation of the observed ultra-high 
energy
cosmic rays (and possibly neutrino oscillations), may introduce an
energy-dependent photon mass that would thus change the speed of light 
at high 
energies \cite{double}. Changes
of the photon speed are also predicted in quantum gravity theories
\cite{QG}. Within these approaches,
 the propagation of light in modified
space-time exhibits a non-trivial
dispersion relation in vacuum, corresponding to Lorentz violation 
through 
an
energy-dependent velocity of light. It was also pointed out \cite{QG}
that one powerful way to probe this possibility may be provided by
some distant astrophysical sources of energetic photons that exhibit
significant and rapid variations in time, such as gamma-ray bursters
(GRBs).
However, in quantum-gravity models, this modification is expected to
be significant at photon energies close to the Planck scale.

In the meantime, in brane-world scenarios \cite{brane}, 
the energy
dependence of the photons' speed can appear at lower energies, since the 
brane
scale is assumed to be much smaller than  Planck's.
 Taking this advantage,
in the present paper we discuss the modification of the velocity of light 
with 
respect 
to the energy of a photon in  the framework of such brane-world scenarios 
setups. 
 In particular, we consider the propagation of photons on a fat 
brane~\cite{fat}, with 
 Standard Model (SM) particles 
localized 
gravitationally 
on the 
brane. Because of the relatively low scale (${\cal O}(1)$~TeV) of the 
massive 
extra-dimensional ``Kaluza--Klein'' 
(KK)  excitations currently under analysis 
\cite{rizzo,pdg,LEP} for the
experimental 
 signatures of 
brane-world scenarios, it is expected that there is
a significant delay of the 
arrival time of lower-energy photons relative to the higher-energy ones, 
propagating 
through large cosmological distances from remote sources. 
 
 Similar effects have been 
proposed in quantum-gravity models \cite{QG} which have recently been 
studied 
in detail in  
\cite{wavegrb,we},  
  where the robust limit on the violation of Lorentz invariance 
has 
been established.
 
 In this paper, we first consider the photon propagation on the brane
 and derive
the dependence of the observed longitudinal velocity on the mass of the KK 
excitations, and then find the relation with  experimental 
measurements  through the light refractive index. The method 
elaborated in \cite{wavegrb,we}, using  
wavelet techniques,
is applied to the available data on GRBs
to estimate the limit on the mass of the KK excitations. 
 This estimate is based on the analysis of the correlation with 
the redshift of
the time-lags between the arrival times of sharp transients in GRB light
curves, observed in higher- and lower-energy bands,
 due to the photon propagation in the expanding brane-world 
Universe.   
 We must stress that
a set of GRBs is studied, and not a single source, since,  as was 
argued in 
\cite{ellisbound,wavegrb,we}, 
 one cannot rely on a single GRB \cite{1sourceGRB},
for which it would be impossible to distinguish between intrinsic time-lag 
at
the origin and a delay induced by the propagation in vacuum. This is 
particularly important
if
the observation of the source is uncertain in a crucial energy band, and 
this 
is why such  an
 estimate may be misleading. 
A large sample of 35 GRBs, with different redshifts and in measured 
different energy bands 
by BATSE, HETE and SWIFT space instruments, is 
systematically analysed. 
   Using the most robust
spectral features of the GRBs light curves, we estimate the 95\% C.L. 
lower
limit on the 
energy scale of the
effective refractive index, and thus on the
inverse thickness of the brane, or on the mass
scale of the KK excitations.

\section{Photon on the brane}

The brane-world idea \cite{brane} provides possible new solutions
to such long-standing problems of particle physics beyond the SM
and cosmology as the hierarchy problem, the smallness of the
cosmological constant, the nature of flavour, the hierarchy of
fermion masses and mixings, etc. The key ingredient of the
brane-world scenarios is that Newtonian gravity and the SM fields
are constrained to be localized on a brane in $n$ extra
dimensions, while the gravitons, and perhaps other SM particles,
are free to propagate in the bulk of extra dimensions. Since the
gravity has the unique feature to be universally coupled with all
kinds of SM fields, the most economic way to trap these fields on
the brane would be to consider models with a mechanism of purely
gravitational localization. Such models, with gravitational
trapping of zero modes of all SM fields, including photons, have
been investigated in~\cite{local}.

In this paper for simplicity in this paper we limit ourselves to a construction
with only one extra dimension, where the matter fields are trapped
by some potential wheel within a fat brane of width $\epsilon =M^{-1}$ along the extra (space-like) 
coordinate $z$. Since the SM gauge fields can only propagate inside the brane, the $\epsilon$
effectively acts as the size of the extra dimension for them.
Therefore, at energies above $M$, the KK excitations (the higher
harmonics of a particle in a box) of the gauge fields can be produced.
The current bound, owing to the lack of capability to excite the
KK modes in different reactions, is close to ${\cal O}(1)$~TeV, as
imposed by LEP \cite{LEP}.

In the following we do not consider any specific case of the trapping
potential on the brane, while just introducing some general
assumptions about the structure of space-time. To be able, in
principle, to localize the Newtonian gravity on the brane,  we
consider the metric to have the standard structure with a warp
factor. In the other words, the 4-dimensional part of the
multidimensional metric depends conformally upon $z$. To avoid
gravitational singularities we also assume that the warp factor is
represented by an even function of $z$. Then one can choose the
following toy-ansatz for the metric
\begin{equation}\label{Ansatz}
ds^2  = e^{-z^2/\epsilon^2}dl^2  - dz^2 ~,
\end{equation}
where
\begin{equation}
dl^2  = g_{\alpha \beta } \left( {x^\nu } \right)dx^\alpha dx^\beta
\end{equation}
is the metric of the 4-dimensional space ($\alpha, \beta,... = 0,
1, 2, 3$) and $e^{-z^2/\epsilon^2}$ is the gravitational warp
factor, which mimics the dependence of the gravitational potential
on the extra coordinate inside the brane. When $z$ goes to 0,
where the brane is assumed to be centred in the transverse
direction, the ansatz (\ref{Ansatz}) describes the 5-dimensional
Minkowski space. As soon as the metric (\ref{Ansatz}) does not
provide gravitational localization of photons on the brane, one
can introduce, in 5-dimensional models, different mechanisms of
photon trapping such as, for example, adding both bulk-\ and
brane-localized mass terms \cite{bulkmass}, or considering models
with higher dimensions \cite{local}. For the present
consideration, the  only  important condition is that SM fields,
placed within a fat brane at different points along the transverse
direction, are exposed to different gravitational potentials, as
follows from (\ref{Ansatz}).
 
It is clear that in our 3-dimensional world the propagation of
light is associated  only with the longitudinal direction inside
the brane. This  means that, once a photon is placed by emission
in any reaction at some distance from $z=0$, it then propagates in
constant gravitational potential. Therefore, it is instructive to
use an interpretation that can be rephrased  as if it were applied
to the light propagation in static brane potential. Namely,
following~\cite{Okun}, one can expect that the frequency $\omega$
of the photon, and hence its energy $E=\omega\hbar$, do not depend
upon the gravitational potential, while the momentum, and thus the
velocity, are expected to be different for photons propagating at
different transverse distances from the centre of the brane. We
shall use below the system of units where $\hbar = 1$, while $c
\neq 1$ in standard units of $c_0$, the speed of light.

A massless photon in five dimensions obeys  the equation
\begin{equation}\label{P}
g^{AB}P_AP_B = 0~,
\end{equation}
where $P_A$ is the 5-momentum and  $g^{AB}$ defines the covariant
components of the metric (\ref{Ansatz}) with capital Latin indexes
running over 0, 1, 2, 3 and  5. The momentum of a photon on the
brane can be represented as the superposition of two components,
namely, one along the longitudinal direction and another one in
the transverse direction relative to the brane. Therefore, the
dispersion relation (\ref{P}) for the {ansatz} (\ref{Ansatz})
takes the form
\begin{equation} \label{v}
\frac{E^2}{v^2} - P^2 - e^{- z^2/\epsilon^2}P_z^2 = 0\, .
\end{equation}
Here $v$ is the velocity of the photon on the brane, and  $P$ and
$P_z$ are the components of the momentum in the directions
parallel and perpendicular to the brane, respectively.

Note that since we assume that photons can only propagate inside
the brane their wave functions are quantized in the potential
wheel of the brane, therefore there is always a mass gap in there
spectrum. The appearance of the transverse component  $P_z$ of the
momentum for the brane observer is equivalent to the occurrence of
the massive KK excitations of the photon. So, there is no already
the massless zero mode in the model and the first quantum level
for the quantized in the brane's potential photon responses to a
mass, which depends on the parameters of the model, for example,
upon the width of the brane $\epsilon$, trapping and gravitational
potentials, etc. This means that the effective {\it longitudinal}
speed $v$ on the brane should be smaller than that in the  bulk,
$c$. Once the transverse component of the total momentum, and
hence the KK excitations, are fixed to $0$, the usual dispersion
relation of the "zero-mass" photon is restored from (\ref{v}):
\begin{equation} \label{c}
\frac{E^2}{c^2}= P^2~.
\end{equation}
Note that the introduction of the effective mass of the photon for
the brane observer, unlike the case where the 4-dimensional
rest-mass term is postulated, does not lead to a loss of gauge
invariance.

Further on, the longitudinal and the transverse components
of the momentum of a particle which sees 5-dimensional
Minkowski space-time should be of the same order:
\begin{equation}
P \sim P_z~.
\end{equation}
However, the transverse component in (\ref{v}) is effectively
suppressed by the gravitational warp factor $e^{-z^2/\epsilon^2}$.
Then, it follows that the effective longitudinal velocity of the
photon is
\begin{equation}\label{vc}
v^2 = \frac{c^2}{1+e^{-z^2/\epsilon^2}} \approx \frac{c^2}{2}\left( 1+
\frac{z^2}{\epsilon^2}\right)~,
\end{equation}
and that it now depends on the position of the photon with respect
to the extra coordinate $z$ inside the brane.

Let us repeat here that our metric (\ref{Ansatz})
represents just a ``toy-ansatz'' and the shape of the warp
factor and the potential, by which the photons and other SM particles are
trapped, does not play a significant role. The important only is that
in the linear approximation we use the
odd $z$-terms do not appear in the expansion 
(\ref{vc}), so it starts from terms of the $z^2$ dependence.
It turns out from (\ref{vc}) that the velocity of a photon at the
centre of the brane ($z=0$) is reduced by a factor $\sqrt{2}$ with
respect to the speed $c$ of light in the bulk and becomes equal to
$c$ only if the photon propagates in the transverse direction at
the distance $z=\epsilon$ from the centre. The latter means that,
for a  photon placed almost out of the brane, the transverse
component of the velocity tends to 0, owing to the trapping caused
by the gravitational potential, the latter increasing from the
centre of the brane toward the bulk.

We can estimate the potential energy $U$ of a particle of mass
$m$, at a distance $z$ from the centre of the source of the static
gravitational field, as
\begin{equation} \label{U}
U \sim maz ~,
\end{equation}
where $a \sim \Gamma^z_{00}$ represents the $z$-component of the
particle gravitational acceleration. The particle with higher
energy could penetrate to a larger distance along the extra
coordinate. Therefore, there should exist also the escape
 energy,
\begin{equation}\label{E}
E_\epsilon \sim Mc^2~,
\end{equation}
where $M \sim 1/(c\,\epsilon)$ is the scale at which the brane,
viewed as a topological defect in higher-dimensional space-time,
was formed. Once the potential energy $U$ of a particle exceeds
$E_\epsilon$, the particle escapes from our world to the
bulk~\cite{Speed}.
 
Then, the energy $E$ of a photon is proportional to the energy of
the particle by which it is emitted. Therefore, as  follows from
(\ref{U}), the characteristic distance at which the emitted photon
gets placed relative to  the centre of the brane is given by
\begin{equation}\label{z}
z \sim E~.
\end{equation}
Now, it is clear from (\ref{vc}) that photons with higher energy
should propagate faster than those with lower energies. Using
(\ref{E}) and (\ref{z}), we can express the transverse coordinate
of the photon $z^2/\epsilon^2$ in (\ref{vc}) via
\begin{equation}
\frac{z^2}{\epsilon^2} \sim \frac{E^2}{E_\epsilon^2}
\sim \frac{E^2}{M^2c^4}~.
\end{equation}
Finally, (\ref{vc}) reads
\begin{equation}\label{speed}
v = \frac{c}{\sqrt{2}} \left( 1+\frac{E^2}{M^2c^4} \right)^{1/2}~.
\end{equation}
The latter means that, because of the influence of the massive KK
excitations on the fat brane, the photons with higher energy
should have higher observable (longitudinal in the brane)
velocity.
 
Let us consider a classical analogy to the propagation of photons as discussed here, which can be 
found as photon propagation
along a waveguide. An ideal waveguide imposes a ``cut-off frequency'' on a propagating
electromagnetic wave based on the geometry of the tube, and will not sustain waves of any lower
frequency. The group velocity in a waveguide is always less than the velocity of light in
vacuum~\cite{jac}. In our consideration a new
 ingradient  such as  gravity is introduced, which changes universally the
transverse component of the photon momentum being responsible for its mass. This
would correspond to the fact that the photons of different energies propagate
along waveguides with different ``cut-off frequencies''.

To relate the result (\ref{speed}) with  experimental
measurements, one generally requires the propagating photon to
have the energy $E$ much smaller than the mass scale $M$ of the KK
excitations, which may be of the order of a TeV, in order to take
brane-world scenarios accessible for the probe on present and
future colliders. This, according to Eq. (\ref{speed}),  implies
the following energy dependence of the velocity $v$ of light: \beq
\label{spped_n} v=c_0\,n(E)~, \eeq where
\begin{equation}
\label{n}
n(E) =1 + \frac{E^2}{8M^2}~
\end{equation}
is the effective vacuum refractive
index\footnote{We use the
relation
$c=\sqrt{2}c_0$ between the bulk and the standard speed of light $c_0$.}.


\section{Light Propagation in the Expanding Brane-World Universe}
To search for the signature of brane-world scenarios,
we compare the propagation of photons with energies much
smaller than $M$, characterizing the divergence of the
vacuum refractive index from unity. A
small
difference between the velocities of two photons with an
energy difference $\Delta E$, emitted simultaneously by a remote
cosmological source, would lead to a time-lag between the arrival
times of the photons.

In what follows we take into account that the propagation 
of photons from a remote object is affected by the expansion of the 
Universe and depends upon the
cosmological model. Present cosmological data motivate the choice of
a spatially-flat Universe: $\Omega_{\rm total} = \Omega_{\Lambda} +
\Omega_M = 1$ with cosmological constant $\Omega_{\Lambda} \simeq
0.7$. The corresponding differential relation between time and
redshift is
\begin{equation} \label{timez}
dt = - \frac{1}{H_0}\frac{dz}{(1 + z)h(z)}~,
\end{equation}
where $H_0$ is the Hubble constant and
\begin{equation} \label{h}
h(z) = \sqrt{\Omega_{\Lambda} + \Omega_M (1 + z)^3}~.
\end{equation}
Thus, a particle of velocity $v$ travels an elementary
distance
\begin{equation} \label{eldist}
vdt=-\frac{1}{H_0}\frac{vdz}{(1+z)h(z)}~,
\end{equation}
giving the following difference between the distances covered by two
particles of velocities differing by $\Delta v$:
\begin{equation} \label{distdiff}
\Delta L = \frac{1}{H_0}\int\limits_0^z\frac{\Delta
vdz}{(1 + z) h(z)}~.
\end{equation}
We can consider two photons travelling with velocities very close to
$c$, whose present-day energies are $E_1$ and $E_2$, where $E_1
> E_2$. At earlier epochs, their energies would have been blue-shifted
by a factor $1+z$. Defining $\Delta E \equiv E_2 - E_1$, we infer
from Eq. (\ref{speed}) that
\begin{equation} \label{Elin}
\Delta v =\frac{\Delta E(E_1+E_2)(1+z)^2}{8M^2}~.
\end{equation}
Inserting the last expressions into (\ref{distdiff}) one finally finds 
that the brane-world 
scenario
induced differences in the arrival times of the two photons of
energy difference $\Delta E$ is
\begin{equation} \label{timedel2}
\Delta t = \frac{\Delta E(E_1+E_2)}{8H_0M^2}\int\limits_0^z
\frac{(1+z)dz}{h(z)}~.
\end{equation}

To look for such a refractive effect induced by the brane-world setup, we 
need a
distant, transient source of photons of different energies,
preferably as high as possible. One may then measure
the differences in the arrival times of sharp transitions in the
signals in
different energy bands. GRBs are at cosmological distances, as
inferred from their redshifts, and exhibit many transient features in
their time series in different energy bands. In comparison, the
observed active galactic nuclei (AGNs) have lower redshifts and broader
time structures in their emissions, but have the advantage of higher
photon energies \cite{AGN}. Observable pulsars  have very 
well defined
time structures in their emissions, but are only at galactic 
distances~\cite{pulsar}. Moreover, the key 
issue in all 
such probes is to distinguish the effects of the vacuum refraction from any intrinsic delay 
in the
emission of photons of different energies by the source. It is obvious
from Eq. \form{timedel2} that the effect of the brane-induced time delay 
should increase 
with the redshift of the source, whereas source
effects would be independent of the redshift in the absence of any
cosmological evolution effects
\cite{ellisbound,wavegrb,we}. Therefore, in order to
disentangle source and propagation effects, it is preferable to use
transient sources of high energy radiation with a broad spread in known 
redshifts $z$. Thus, 
one of
the
most model-independent ways to probe the time-lags that might arise from 
the brane-world 
setup  is to use the GRBs with known redshifts, which range up to
$z
\sim 6$.

In the present paper, we study the brane-world-induced light refraction in 
vacuum, compiling 
the results from~\cite{we} on time-lag measurements for a 
sample of 35 GRBs with known
redshifts, including 9 GRBs detected by the Burst And Transient
Source Experiment (BATSE) aboard the Compton Gamma Ray Observatory
(CGRO),
15 detected by the High Energy Transient Explorer (HETE)
satellite and 11
detected by the SWIFT satellite. The GRBs currently in use are listed in Table 
1 of~\cite{we}, together with their redshifts and the
time-lags; the latter were extracted from their light curves by using the 
special wavelet 
techniques pioneered 
in~\cite{wavegrb} and developed further in~\cite{we}. The gamma ray light 
curves in~\cite{we}  
 are from
BATSE~\cite{cgro},
HETE~\cite{hete} and SWIFT~\cite{swift} public archives, and  the 
information on the redshifts is
from~\cite{z_info}. \label{sec_prop}  

\section{GRB Constraint on Mass of ``Kaluza--Klein'' \\  Excitations on a 
Fat Brane}

As just discussed above, 
Eq. \form{timedel2} may
be accompanied by {\it a priori} unknown intrinsic
energy-dependent time-lags, caused by
unknown properties of the sources. To take this into 
account we use the fit $\Delta t(z)$ of the measured time-lags where we  
include a 
term
$b_{\rm sf}$ specified in the rest frame of the source, so that the
resulting observed arrival time delays $\Delta t_{\rm obs}$ are fitted by 
two
contributions:
\beq
\label{lv+rf}
\Delta t_{\rm obs}=\Delta t_{\rm KK}+b_{\rm sf}(1+z),
\eeq
one, $\Delta t_{\rm KK}$, reflecting the possible effects of the 
brane-world setup, and another one representing  
intrinsic source effects.
Rescaling \form{lv+rf} by a factor $(1+z)$, we arrive at a
simple linear fitting function
\beq
\label{lv+rf+scale}
\frac{\Delta t_{\rm obs}}{1+z}=a_{\rm KK}K+b_{\rm sf},
\eeq
where
\beq
\label{K1}
K \equiv \frac{1}{1+z}\int\limits_0^z\frac{(1+z)dz}{h(z)}
\eeq
is a non-linear function of the redshift $z$  related to the 
measure
of the
cosmic distance in \form{timedel2}, and the 
slope coefficient 
\beq
\label{akk}
a_{\rm KK}=H_0^{-1}\frac{\Delta
E(E_1+E_2)}{8M^2} 
\eeq
 in $K$ is connected to the mass scale of KK excitations. 
 The energy difference in~\form{timedel2} is taken in a way that the
negative slope parameter $a_{\rm KK}$ in the fit would correspond to the
refractive index~\form{n} predicted in brane-world scenarios.  The GRB 
data  \cite{we} 
used here allows looking for spectral time-lags in the
light curves recorded at the energy $E_1=400$~keV, relative to those at
lowest $E_2=30$~keV energy.

Before proceeding further and finding the estimate on the mass of 
the 
KK excitations,  we want to underline
that the procedure applied here follows 
that of \cite{wavegrb,we} and 
allows 
 arriving at firm conclusions. 
Let us dwell on 
 the main statements of the procedure, given in detail in 
\cite{we}: 

\begin{itemize}
\item
To disentangle the most significant variations of time profiles,
 the GRB data light curves undergo a complex handling procedure, using
discrete and continuous wavelet transforms. 
In this way the most
singular, or ``genuine'' sharp points in the higher- and the lower-energy
spectral band counterparts are determined to be used in 
\form{lv+rf+scale}. 
 \item The obtained points are carefully checked for statistical 
stability. 
 In each energy band, artificial noise 
has been generated and added to the original light curves,
and the above wavelet search has been repeated  a couple
of thousand times per band, to provide reasonable convergence
of the noise iteration process.
 \item The contamination of the real data with artificial noise is applied
to allow statistically significant estimates of the errors in the
determination of the signals positions.
 \item Particular attention has been devoted to figuring out how
sensitive the fit results are to the unknown uncertainties due 
to the
intrinsic properties of the sources we use.  Several procedures of
weighting the estimated time-lags and their (systematic) errors have been 
considered. 
It was revealed that the most conservative way to estimate the
time-lags and the errors is to compare the arrival time of those photons
coming from a given GRB when the highest parts of emission are
progressing while a universal stochastic spread in the intrinsic 
time-lag at 
the source is allowed. 

\end{itemize}

{\small
\begin{figure}[t]
\centerline{\psfig{file=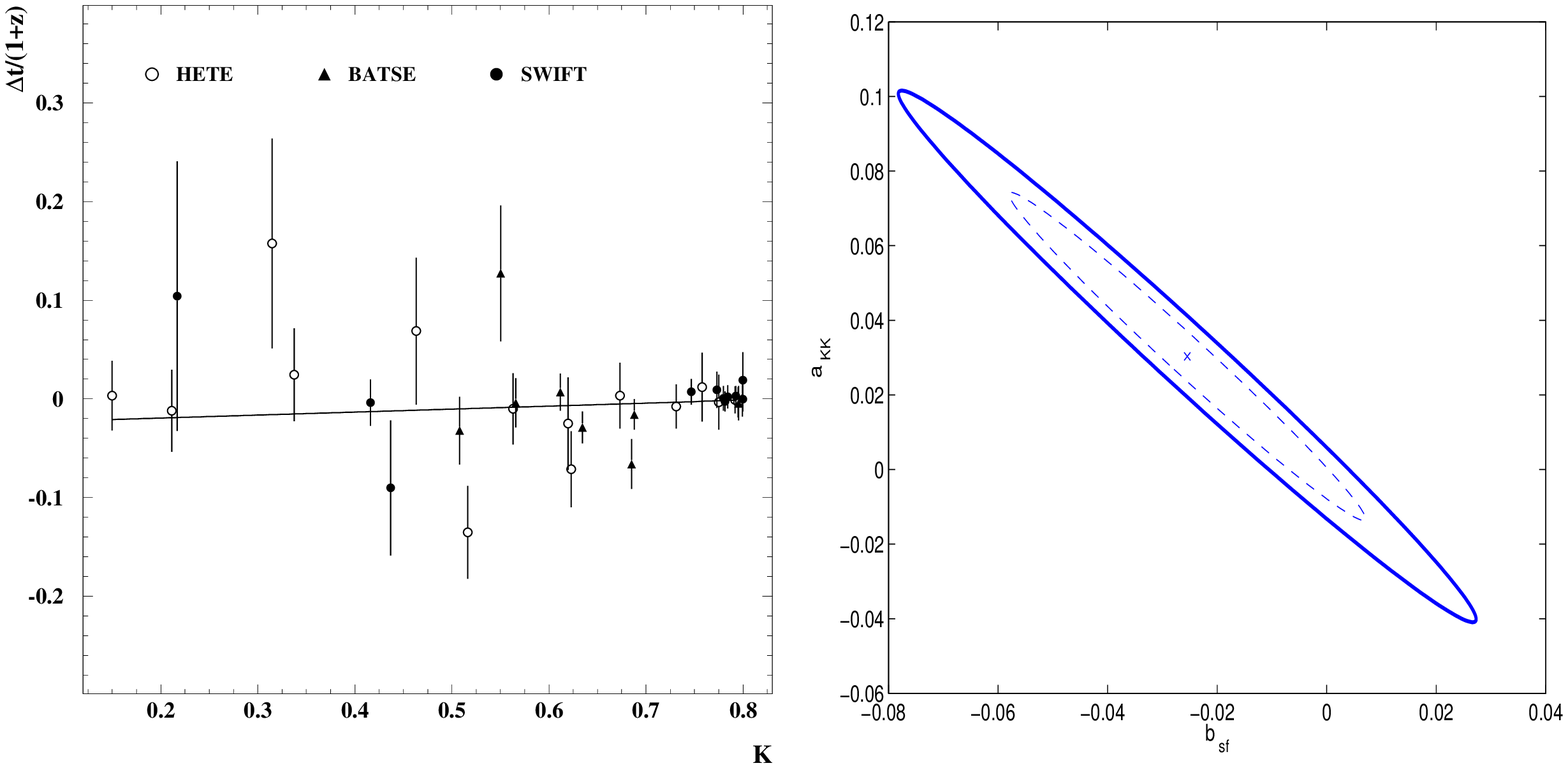,height=7cm,width=16cm}}
\vspace*{8pt}
\caption{
 \small
{\bf Left panel}: 
 {The rescaled spectral time-lags at highest pulses between the arrival
times of pairs of genuine sharp features, wavelet-extracted \cite{we} from
the light curves of the full set of 35 GRBs, with measured redshifts
observed at time resolutions of 64~ms (BATSE, SWIFT) and 164~ms (HETE),
and a linear fit \form{fir64l''} to these data with ${\chi^2/{\rm
d.o.f.}=32.8/33}$.
 The data are normalized to the difference between the energies
$E_1=400$~keV and $E_2=30$~keV of the third and first HETE spectral bands.
 The errors in the redshifts and hence in $K$ are negligible: the errors
in the time-lags are estimated by the wavelet analysis \cite{we} and
increased by 35~ms, modelling a possible stochastic spread at the sources
(see text).}
 {\bf Right panel}: {The error ellipse in
the slope-intercept plane for the fit~\form{fir64l''}. The 68\% and 95\%
confidence-level contours are represented by the dashed and solid lines,
respectively.}} 
 \label{regr1} 
 \end{figure}
}

As seen, \form{lv+rf+scale} 
depends 
linearly 
on $K$, and therefore, in
order to probe the energy dependence of the velocity of
light that
might be induced by the brane-world,
we
perform a linear regression analysis of the rescaled by $(1+z)$ time-lags
(\ref{timedel2}) of the data studied here. These data, representing a set
of measurements from 35 GRBs 
and compiled from Table 1  
of~\cite{we}, as described above,  are shown here
in the left panel of Fig.~\ref{regr1}
 as a function of $K$ (\ref{K1}).\footnote{
 Since the data from the HETE and SWIFT instruments are made available in
 slightly different energy bands, we have rescaled the time-lags and 
 errors from Table 1 of~\cite{we} by the ratios of the energy differences 
 of the 
 HETE and SWIFT data relative to those of the BATSE instrument.}

The result
of the straight-line fit~\form{lv+rf+scale} 
leads to
\beq
\label{fir64l''}
\left(\frac{\Delta t_{\rm obs}}{1+z}\right)\;
= (0.030 \pm 0.029)\, K\; -\; (0.025 \pm 0.021),
\eeq    
 and 
 is also shown in the left panel of Fig.~\ref{regr1}.
To take 
systematics 
into 
account, 
we allow for a universal spread in the intrinsic  
time-lag at the
source  by adding in quadrature, for all the GRBs,
a universal source error whose normalization is then fixed so that
$\chi^2/{\rm d.o.f.}=1$. The corresponding universal source error
is estimated in this way to be 35~ms,\footnote{This universal source 
error is well within the resolution 
of all the instruments whose data are used for the analysis.}  which has 
been added to the errors of 
time-lags from \cite{we}. The
fit \form{fir64l''} yields a
positive value of the slope parameter $a_{\rm KK}$ at the 1$\sigma$
level, while the  slope in the model we  wish to explore must be 
negative semi-definite. Moreover, 
the slope and intrinsic time-lag parameters are 
highly correlated, as seen in the right panel of Fig.~\ref{regr1}. This 
certainly cannot be considered as 
evidence for the absence of vacuum refractive index induced by brane-world 
scenarios and will be used to 
quote a limit on the mass scale $M$ of the KK excitation mode, which is 
supposed to be one of the main  
parameters to be addressed  in studies of the brane-world 
phenomenology at future 
colliders  \cite{lhc-ilc-physreport}. 

To estimate the lower limit of the energy required to excite the zero KK 
mode on a fat brane, we use the 
marginal distribution of the slope
parameter in the fit~\form{fir64l''}, as obtained by integrating over the
intercept parameter. Taking into account the correlation matrix as
described in~\cite{barlow_book}, we rescale by a factor $\sqrt{1-\rho^2}$
the Gaussian-like shape of this marginal distribution, where $\rho$ is the
correlation coefficient of the bivariate slope-intercept distribution. The
mean value is still unchanged at~$a_{\rm KK}=0.030$,
whereas the variance (defined as the width at  half-maximum)  
is $\sigma_{a_{\rm KK}}=0.025$, which is still 
1.2$\sigma$ above zero. 

We quote a
limit on $M$ in the
Bayesian manner
proposed
in~\cite{neurtinolimit}, where the confidence range was constructed for
a Gaussian distribution with positive mean, which is physically 
constrained
to
be negative. 
For the measured positive mean of
the marginalized slope distribution at
1.2$\sigma$ above zero, we calculate the 95\% confidence limit
on
the mass scale of the KK excitations,  assuming a
random variable obeying Gaussian
statistics with a boundary at the origin.
 The corresponding upper limit on the negative value of the slope 
parameter
is $a_{\rm KK}^{\rm min}=-0.024$.\footnote{This value is obtained by
multiplying the upper edge of the 95\% confidence interval $(-0.97)$ with
$\sigma_{a_{\rm KK}}=0.025$ listed in Table X of~\cite{neurtinolimit}, at 
the
line corresponding to $x_0=-1.2$.} 
 Inserting $a_{\rm KK}^{\rm min}$ into
\form{akk} and solving it for $M$, one gets
 \beq
\label{scaleM}
M=\sqrt{H_0^{-1}\frac{\Delta
E(E_1+E_2)}{8\,a_{\rm KK}}},
\eeq
 which, with the energies $E_1=400$~keV and $E_2=30$~keV one 
deals with, leads to the lower limit
\beq
\label{order_stat}
M \ge 620 \; {\rm TeV}
\eeq
on the masses of the KK excitations, due to the light refraction 
in 
vacuum of the brane-world Universe.
 A similar result is  obtained from using the likelihood method as 
an alternative to this
Bayesian approach, as well as with another way of scaling the errors.
This is similar to what has also been obtained in \cite{we}, 
where details on 
the likelihood
approach and additional way of error scaling can be found.

\section{Conclusions}

In summary, we have investigated possible non-trivial refractive
properties of vacuum induced by models with a fat brane-world embedded 
into
a higher-dimensional space-time. This feature can appear for those setups
where the SM particles are localized gravitationally on the brane. 
 Then, as shown, the energy of the particle emitting a photon defines the 
frequency of the latter, and thus its momentum transverse component 
 contributing as the massive KK  excitations in the  
spatial extra dimension.  We have
derived the effective vacuum refractive index for photons propagating on a
fat brane as a function of the mass of the KK  excitations. 

To search 
for 
 observable
signatures of brane-world scenarios, we have related the 
 obtained 
refractive index 
formula 
 with the arrival time-lags  of two photons 
with different 
energies emitted simultaneously by a remote cosmological source.  
The expression found has been 
applied to  recent compilation results  \cite{we}, obtained with 
wavelet 
techniques  exploiting  data from 35 GRBs
with known redshifts,
to estimate
correlation with redshift of the time-lags between the arrival times of 
sharp
transients in GRB light curves observed in higher- and lower-energy 
bands.

The analysis does not show any significant correlation of the measured 
time-lags
with the cosmological redshift to indicate any 
deviation of the
vacuum refractive index from unity. This fact allows us to establish 95\%
C.L. lower limit on the inverse thickness of the brane and hence on the
mass scale of the KK excitations at the level of 620~TeV. 
 The constraint 
obtained
is at least an order of magnitude higher than the energy achievable at
future high energy colliders and is of great interest in the ongoing
 discussion on experimental 
scale probes of the considered 
class of brane-world scenarios.

\bigskip

\noindent
{\bf \Large Acknowledgements}
\medskip

We thank John Ellis and Massimo Giovannini for useful discussions. 
M.G. acknowledges the hospitality of CERN TH Division, where this work 
has been started. AS thanks ITP at EPFL (Switzerland) and MIUR grant under the 
project 
 PRIN 2004 ``Astroparticle Physics'' at Physics Department of University 
L'Aquila (Italy) for partial support. 

\vspace*{-0.4cm}


\end{document}